%% file: build-cache.tex
\pdfoutput=1
\RequirePackage{amssymb}
\RequirePackage[l2tabu,orthodox,abort]{nag}
\PassOptionsToPackage{final,stretch=10,spacing=true}{microtype}
\PassOptionsToPackage{hyphens,obeyspaces}{url}
\documentclass[sigconf,review,natbib=false]{acmart}

\usepackage[utf8]{inputenc}
\usepackage[datamodel=acmdatamodel,style=acmnumeric,maxnames=3]{biblatex}
\usepackage{booktabs}
\usepackage{colortbl}
\usepackage{enumitem}
\usepackage{graphicx}
\usepackage{listings}
\usepackage{siunitx}
\usepackage{tikz}

\usepackage[repro]{reidpr}
\pdfid{2554cb3a-8bf3-4f65-a99c-2844d00e7bf0}

\graphicspath{{figures/}}
\newcommand{\treescale}{0.686}

\definecolor{c:fastclip}{rgb}{0.1843, 0.4726, 0.7116}
\definecolor{c:equal}{rgb}{0.9657, 0.9672, 0.9680}
\definecolor{c:slowclip}{rgb}{0.7561, 0.2103, 0.2235}

\setlist[description]{leftmargin=\parindent,listparindent=\parindent}
\setlist[enumerate]{listparindent=\parindent}
\def\DefaultCutFileName{\def\CommentCutFile{\jobname.cut}}
\DefaultCutFileName

\makeatletter
\protected\def\abx@missing#1{%
  \mbox{\reset@font\textcolor{\fixmecolor}{#1}}}
\makeatother
\newcommand{\ch}{\emph{ch}}
\newcommand{\chd}{\emph{ch–}}
\newcommand{\chgc}{\emph{ch compacted}}
\newcommand{\dko}{\emph{dko}}
\newcommand{\pmo}{\emph{pmo}}
\newcommand{\id}{state ID}
\newcommand{\Id}{State ID}
\newcommand{\image}[1]{\code{#1}}

\lstnewenvironment{dfsmall}[2]
  {\lstset{title={\raggedright\small\texttt{#1}},
           linewidth=#2,
           numbers=none,
           frame=single,
           framerule=0.2pt}}
  {}

\makeatletter
\def\fps@figure{tbp}
\def\fps@table{tbp}
\makeatother

\title{Charliecloud’s layer-free, Git-based container build cache}
\author{Reid Priedhorsky}
\orcid{0000-0002-5348-0330}
\email{reidpr@lanl.gov}

\author{Jordan Ogas}
\author{Claude H.\ (Rusty) Davis IV}
\author{Z.\ Noah Hounshel}
\additionalaffiliation{%
  \institution{University of North Carolina Wilmington}
  \city{Wilmington}
  \state{North Carolina}
  \country{USA}
}

\author{Ashlyn Lee}
\additionalaffiliation{%
  \institution{Colorado State University}
  \city{Fort Collins}
  \state{Colorado}
  \country{USA}
}

\author{Benjamin Stormer}
\additionalaffiliation{%
  \institution{University of Texas at Austin}
  \city{Austin}
  \state{Texas}
  \country{USA}
}

\author{R.\ Shane Goff}
\affiliation{%
  \department{High Performance Computing Division}
  \institution{Los Alamos National Laboratory}
  \city{Los Alamos}
  \state{New Mexico}
  \country{USA}
}

\acmYear{2023}
\setcopyright{rightsretained}

\acmConference{Pre-print}{please cite the version of record when available}{thank you}
\acmISBN{}
\acmDOI{}
\makeatletter
\def\ACM@linecountL{}
\def\ACM@linecountR{}
\makeatother

\begin{abstract}

  A popular approach to deploying scientific applications in high performance computing (HPC) is \vocab{Linux containers}, which package an application and all its dependencies as a single unit.
  This \vocab{image} is built by interpreting instructions in a machine-readable recipe, which is faster with a \vocab{build cache} that stores instruction results for re-use. The standard approach (used e.g.\ by Docker and Podman) is a many-layered union filesystem, encoding differences between layers as tar archives.

  We describe a new approach, implemented in Charliecloud: store changing images in a Git repository.
  Our experiments show this performs similarly to layered caches on both build time and disk usage, with a considerable advantage for many-instruction recipes. Our approach also has structural advantages: better diff format, lower cache overhead, and better file de-duplication.
  These results show that a Git-based cache for layer-free container implementations is not only possible but may outperform the layered approach on important dimensions.
\end{abstract}

\begin{document}
\maketitle

\setlength{\fboxsep}{0pt}
\fbox{\parbox[c][20ex]{\columnwidth}{\centering\itshape
    Pre-print dated 2023-08-31.\\
    Please cite the version of record when available.
  }}

\section{Introduction}

\vocab{Linux containers} is a technology for packaging software into \vocab{images} that contain an application along with its complete software stack. This approach provides simplified dependency management, reliable provenance, straightforward archival of complete environments, and improved portability. It is becoming increasingly popular in HPC.

Container images are built using a recipe such as a \vocab{Dockerfile}, which gives a sequence of container operations and arbitrary commands whose execution transforms the empty root image through a sequence of intermediate states into the desired image. A key feature of image build tools is the \vocab{build cache}, which speeds build times by reusing already computed image states. Existing tools such as Docker and Podman implement this cache with a layered (union) filesystem such as OverlayFS~\cite{brown2023overlay} or FUSE-OverlayFS~\cite{scrivano2023fuseoverlayfs} and tar archives to represent the content of each layer; this approach is standardized by the Open Container Initiative (OCI)~\cite{oci2023image}. While effective, the layered cache has drawbacks in three critical areas:

\begin{enumerate}

\item \inhead{Diff format.} The tar format is poorly standardized and not designed for diffs~\cite{sarai2019road}. Notably, tar cannot represent file deletion. The workaround used for OCI layers is specially named \vocab{whiteout} files, which means the tar archives cannot be unpacked by standard UNIX tools and require special container-specific processing.

\item
  \inhead{Cache overhead.} Each time a Dockerfile instruction is started, a new overlay filesystem is mounted atop the existing layer stack. File metadata operations in the instruction then start at the top layer and descend the stack until the layer containing the desired file is reached. The cost of these operations is therefore proportional to the number of layers, i.e., the number of instructions between the empty root image and the instruction being executed. This results in a best practice of large, complex instructions to minimize their number~\cite{docker2023best}, which can conflict with simpler, more numerous instructions the user might prefer.

\item \inhead{De-duplication.} Identical files on layers with an ancestry relationship (i.e., instruction $A$ precedes $B$ in a build) are stored only once. However, identical files on layers without this relationship are stored multiple times. For example, if instructions $B$ and $B'$ both follow $A$ — perhaps because $B$ was modified and the image rebuilt — then any files created by both $B$ and $B'$ will be stored twice~\cite{sarai2019road}.

  Also, similar files are never de-duplicated, regardless of ancestry~\cite{sarai2019road}. For example, if instruction $A$ creates a file and subsequently instruction $B$ modifies a single bit in that file, both versions are stored in their entirety.

% This paragraph seems to not be true.
%
%  Further, filesystem-based de-duplication is of little use in this context. First, layers are stored as compressed tar archives, so trivial differences such as small changes in a file or file ordering cause cascading differences in the rest of the archive. The \code{--rsyncable} option for GNU tar might help with this~\cite{tridgell1999efficient,warnier2005rsyncable}, but the golang gzip library used by Docker and Podman does not support it~\cite{golang2023gzip}.

\end{enumerate}

Charliecloud — LANL’s lightweight, HPC-focused, open source, fully unprivileged container implementation — recently introduced a new approach: cached files are stored in a local Git repository, with large files optionally stored separately. This addresses the three drawbacks: (1)~Git is purpose-built to store changing directory trees, (2)~cache overhead is imposed only at instruction commit time, and (3)~Git de-duplicates both identical and similar files. Also, is based on an extremely widely used tool that enjoys development support from well-resourced actors, in particular on scaling~\cite{blau2022git238,stolee2020scalar}.

This paper details the motivation and design of Charliecloud’s build cache and reports performance experiments showing the Git-based approach performs well in both build time and disk space.

\section{Cache design and operation}
\label{sec:cachedesign}

This section explains how the cache works and why.

\subsection{Background}

\begin{figure}
\begin{lstlisting}[firstline=3]
FROM alpine:3.16
RUN apk add python3
COPY randomfiles /

RUN mkdir /a && mkdir /b
RUN /randomfiles /a 256 512 16384
RUN /randomfiles /b 256 512 16384  #WARM#
\end{lstlisting}
  \caption{Dockerfile that creates \qty{4}{GiB} of random data in $\mathbf{2^{18}}$ small (\qty{16}{KiB}) files, our \protect\image{megafiles} experiment image. Starting from the Alpine Linux 3.16 official image (line~1), which was itself produced using a different Dockerfile starting from the empty image~\cite{copa2023dockeralpine}, this recipe installs a Python interpreter (line~2) and copies our script \protect\code{randomfiles} from the build host (line~3). It then makes two directories (line~5) and creates the files in two steps of $\mathbf{2^{17} = 256{}\times{512}}$ files each (lines~6 and~7).}
  \label{fig:dockerfile}
\end{figure}

Image builders such as Docker, Podman, and Charliecloud’s \code{ch-image} execute a sequence of instructions that progresses the image from the empty starting state through a sequence of intermediate states to the final target state. An image state comprises all the files in the image, their metadata, and some metadata about the container itself (e.g., environment variables). Each instruction combined with the previous state produces a new state. The de facto standard for encoding these instructions is a text format called a \vocab{Dockerfile}~\cite{docker2023dockerfile}; Figure~\ref{fig:dockerfile} shows one Dockerfile used in our performance experiments (detailed below in §\ref{sec:performance}).

Caching is a very well established technique for improving performance. The basic approach is to store computation results, and then if a result is needed again, retrieve it from cache rather than re-computing it. In the case of Dockerfiles, many instructions are expensive to compute (e.g., \code{RUN wget https://slow.example.com/bigfile}), motivating the use of caching to store/retrieve states instead of re-computing them.

\subsection{Design priorities}

Charliecloud’s build cache design priorities are, in descending order of importance:

\begin{enumerate}

\item \inhead{Correctness.} The cache should do what is actually needed and have few bugs.

\item \inhead{Use clarity.} Users should be able to reason correctly about what the cache will do as they use Charliecloud, with minimal documentation reading.

\item \inhead{Implementation clarity.} Developers should be able to understand how the cache works and how to modify its code without undue effort, though this will include design documentation such at this paper.

\item \inhead{Time efficiency.} The cache should be fast.

\item \inhead{Space efficiency.} The cache should use little disk space.

\end{enumerate}

As for any software project, this order is context-sensitive and approximate, but we depend on it for guiding principles.

\subsection{No layers}

Like Singularity/Apptainer~\cite{kurtzer2017singularity,qwertyjack2020better}, Charliecloud is a layer-free container implementation. It has no internal notion of layers\footnote{Charliecloud does use layers when interacting with external resources such as container registries.} and operates with whatever filesystem  (typically flat) is provided. For its build cache, Charliecloud stores each image state as Dockerfile instructions transform the image from the empty base state to the final state. This corresponds to the \vocab{version-oriented} view of version control systems (VCS) used for software development~\cite{lie1989change}.

The more common, OCI-standardized, view is \vocab{change-oriented}. Each instruction creates a new \vocab{layer}. When the image is live (being built or running an application), the stack of layers is represented as a stack of layered filesystems (e.g., OverlayFS). Each layer can then be serialized into tarball containing files new and changed by that layer, along with whiteouts to represent deletions~\cite{oci2023image}: that is, OCI tarballs are simply another diff format.

These views are equivalent. Given two states, one can compute the diff between them, and format it as an OCI tarball if needed. Given a starting state — perhaps the empty image — and a sequence of layers, one can apply these diffs to compute the resulting state. Charliecloud uses this equivalence to communicate with OCI resources such as image registries.

\subsection{False assumptions}
\label{sec:assumptions}

All container build caches make two key assumptions:
\begin{enumerate}

\item \label{ass:retrieval} Retrieving image state is always cheaper than executing the corresponding instruction(s).

\item \label{ass:dependencies} State depends only on visible input and ancestry, i.e., the previous state plus the new instruction.

\end{enumerate}
These assumptions are false. For example, \code{RUN true} is likely faster to execute than retrieve from cache, violating Assumption~\ref{ass:retrieval}, and \code{RUN date > foo} also depends on \emph{invisible} input (the current time), violating Assumption~\ref{ass:dependencies}. Charliecloud mitigates this problem in three ways:

\begin{enumerate}

\item Ignore it.\footnote{This is the authors’ favorite solution to any computer-related problem and in fact deliberately first in the list.} While the assumptions are not strictly true, they are usually “true enough”.

\item Retrieve minimum state needed. Because a state depends on its ancestors, one miss will cause all subsequent instructions to also miss, so any build is a sequence of zero or more cache hits followed by zero or more cache misses. Charliecloud retrieves only the state of the last cache hit.

\item Make the build cache optional. This lets the user use their greater knowledge of the build to deliberately re-execute when needed, either by re-executing all instructions after \code{FROM} (\code{ch-image --rebuild}) or disabling the cache entirely (\code{ch-image --no-cache}).\footnote{An alternative we considered but rejected is separate toggles for build cache reading and writing, by making  no-read convert all reads into misses and no-write convert all writes to no-ops. While this principle is fairly simple, it yields difficult corner cases. For example, with no-read, \code{FROM} cannot copy the base image from cache, so re-building an image will yield a branch starting from the root rather than the base image, even though the build started from the base image. Another corner case is to build normally, then build no-write, then build normally again; we worked out what would happen but it was so confusing that we immediately forgot.}

\end{enumerate}

Additional future mitigations may include:

\begin{itemize}

\item Dockerfile notation to mark an instruction non-cacheable.

\item Finer-grained control of the cache; e.g., a CLI argument that rebuild mode starts on a given Dockerfile line.

\end{itemize}

We next turn to how the cache works.

\subsection{Cache operation by example}

Because each state is assumed to be a function of (1)~exactly one instruction with its visible input and (2)~exactly one \vocab{parent state}, image states (and their corresponding cache entries) form a tree. Each node is identified by its \vocab{\id}, which is a digest of its parent’s \id{}, its instruction text, and that instruction’s visible input (see §\ref{sec:stateid} below for details). This section demonstrates how that tree works by detailing several examples. Later, we detail how the tree is stored in Git.

\subsubsection{Empty cache}

\begin{figure}
  \centering
  \includegraphics[scale=\treescale]{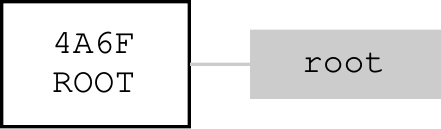}
  \caption{Empty build cache tree upon initialization. In this and related figures below, the first line of each node (white boxes) is the abbreviated \vocab{\id} identifying the state (here \protect\code{4A6F}), and the second line is the instruction that created it (here \protect\code{ROOT}). Branch labels (image names) are gray boxes (here \protect\code{root}). These figures are debug output from Charliecloud.}
  \label{fig:empty}
\end{figure}

Figure~\ref{fig:empty} shows the cache after initialization, containing only the empty root image. This is the one node \code{4A6F} created by the pseudo-instruction \code{ROOT}, which is labeled\footnote{Charliecloud does use other types of labels internally, omitted here for clarity. Readers interested in that level of detail can consult the Charliecloud source code.} \code{root} to indicate that it corresponds to the image named\footnote{Unlike Docker and Podman, Charliecloud has image \vocab{names} rather than image \vocab{tags}. That is, \code{ch-image build -t foo} creates an image named \code{foo}, not an image tagged \code{foo}. This is to reduce confusion related to the highly-overloaded term \vocab{tag}.}  \code{root}. (The debug output in the figures uses 16-bit abbreviated \id{}s for clarity, but they are really 128 bits.) Because the tree root has no parent and no instruction, we use an arbitrary constant \id{}.

\subsubsection{Pull}

\begin{figure}
  \centering
  \includegraphics[scale=\treescale]{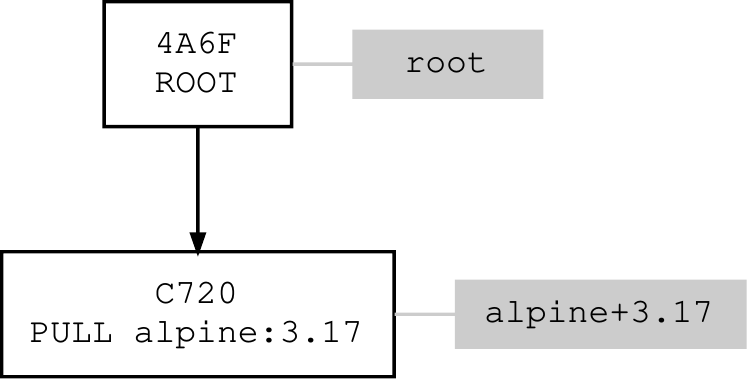}
  \caption{Cache after pulling image \protect\code{alpine:3.17}.}
  \label{fig:initialpull}
\end{figure}

Figure~\ref{fig:initialpull} shows the cache after \code{ch-image pull alpine:3.17}. It now contains one additional node: its \id{} \code{C720} is the digest of the parent ID (\code{4A6F}), the instruction text (\code{PULL alpine:3.17}, another pseudo-instruction), and the visible input, which is the manifest obtained from the image repository during the pull process~\cite{oci2023distribution}. The image name (\code{alpine:3.17}) is encoded in the tree with plus replacing colon to meet Git branch name requirements.

If we were to \code{ch-image pull} again, Charliecloud would download \code{alpine:3.17}’s manifest (again), and using that compute the \id{}. If this \id{} is already in the cache (i.e., the registry image is unchanged), that is a cache hit and the pull is aborted. If it is not found, that is a cache miss and the pull operation proceeds, creating a new child node of the root. In either case, the image label points to the node corresponding to the latest manifest; if that required moving it, then the old node remains but is unlabeled.

\subsubsection{Build a Dockerfile}

\begin{figure}
  \centering
  \begin{tikzpicture}
    \node at (0,0) {
      \includegraphics[scale=\treescale]{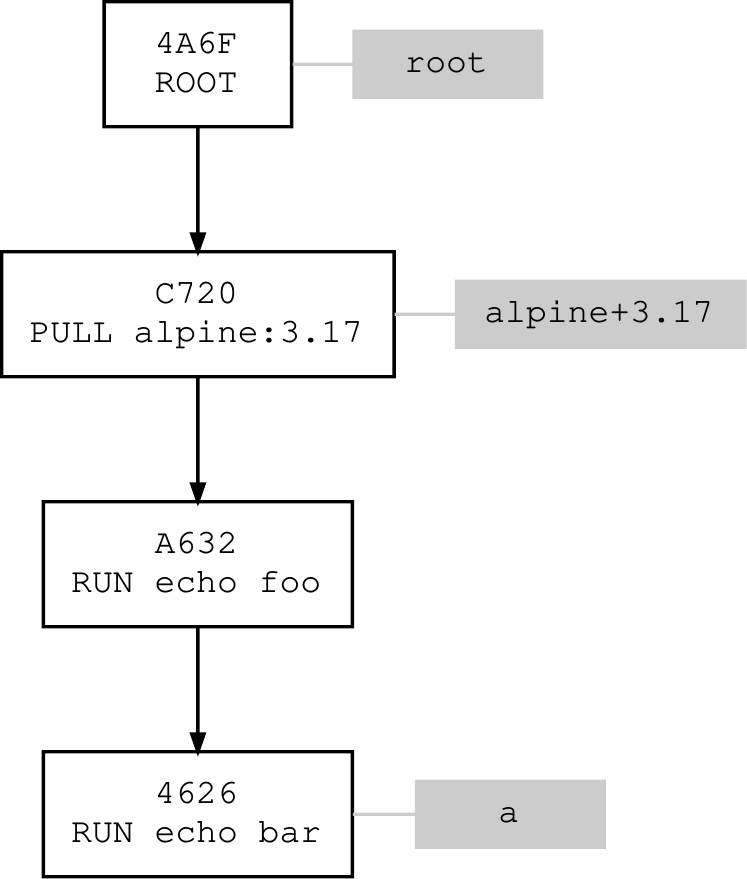}
    };
    \node at (60pt,-18pt) {
\begin{dfsmall}{a.df}{16.5ex}
FROM alpine:3.17
RUN echo foo
RUN echo bar
\end{dfsmall}
    };
  \end{tikzpicture}
  \caption{Cache after building \protect\code{a.df}.}
  \label{fig:builda}
\end{figure}

Figure~\ref{fig:builda} shows the cache after building a simple Dockerfile to create image \code{a}. The first instruction \code{FROM alpine:3.17} ensures there is a branch labeled \code{alpine+3.17}. If this base image had been previously pulled, there is nothing to do, i.e., we start with the tree in Figure~\ref{fig:initialpull};\footnote{Like Docker and Podman, Charliecloud’s \code{FROM} does not check if the base image is up to date, while manual pull does.} if not, we pull it to create that tree. Either results in a node for \code{PULL} with \id{} \code{C720}.

We next compute \code{RUN echo foo}’s \id{} \code{A632}, which is the digest of the parent’s \id{} (\code{C720}) and the instruction text. (\code{RUN} instructions have no other input.) This state is not in the tree as of Figure~\ref{fig:initialpull}, so the instruction is a \vocab{cache miss} and must be executed. We check out the state of the last \vocab{cache hit} (again \code{C720}) into a work directory, execute \code{echo foo} there in a Charliecloud container, and then commit the result into the cache at node \code{A632}.

We know \code{RUN echo bar} is a cache miss because we already had a miss, and we can re-use the work directory from the previous instruction. We execute a containerized \code{echo bar} and check in the root filesystem at \code{4626}. This is the last instruction, so we label that branch tip node as image \code{a}.

Note that there is no cache node for \code{FROM}. This is because \code{FROM} is an unusual instruction that does not \emph{do} anything to the image. Rather, it pulls the base image if needed and links the in-progress image to the base image.

\subsubsection{Build a derived Dockerfile}

\begin{figure}
  \centering
  \begin{tikzpicture}
    \node at (0,0) {
      \includegraphics[scale=\treescale]{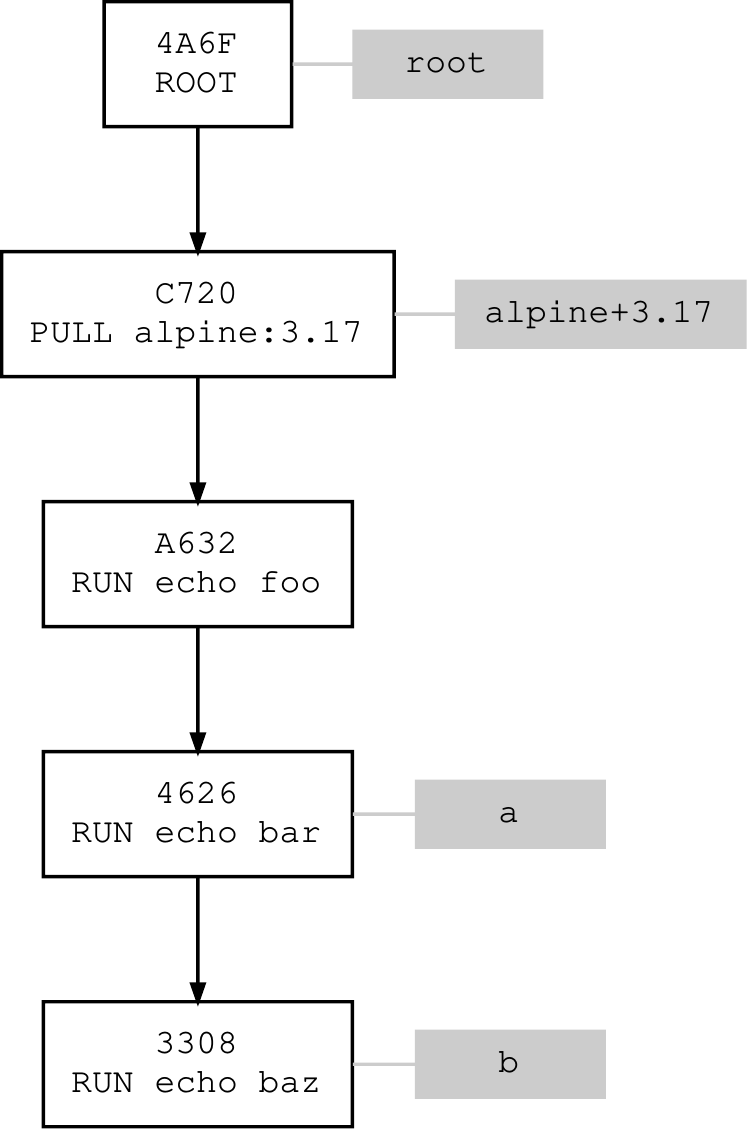}
    };
    \node at (60pt,6pt) {
\begin{dfsmall}{b.df}{12.5ex}
FROM a
RUN echo baz
\end{dfsmall}
    };
  \end{tikzpicture}
  \caption{Cache after building \protect\code{a.df}, then \protect\code{b.df}.}
  \label{fig:buildb}
\end{figure}

Figure~\ref{fig:buildb} shows the cache after building a Dockerfile based on \code{a}, creating \code{b}. Here, it is necessary to build \code{a} first, because if \code{a} isn’t cached, the build will try to pull it and fail. \code{RUN echo baz} is a cache miss, executed, and stored by the same procedure as the previous section.

\subsubsection{Build a related Dockerfile}

\begin{figure}
  \centering
  \begin{tikzpicture}
    \node at (0,0) {
      \includegraphics[scale=\treescale]{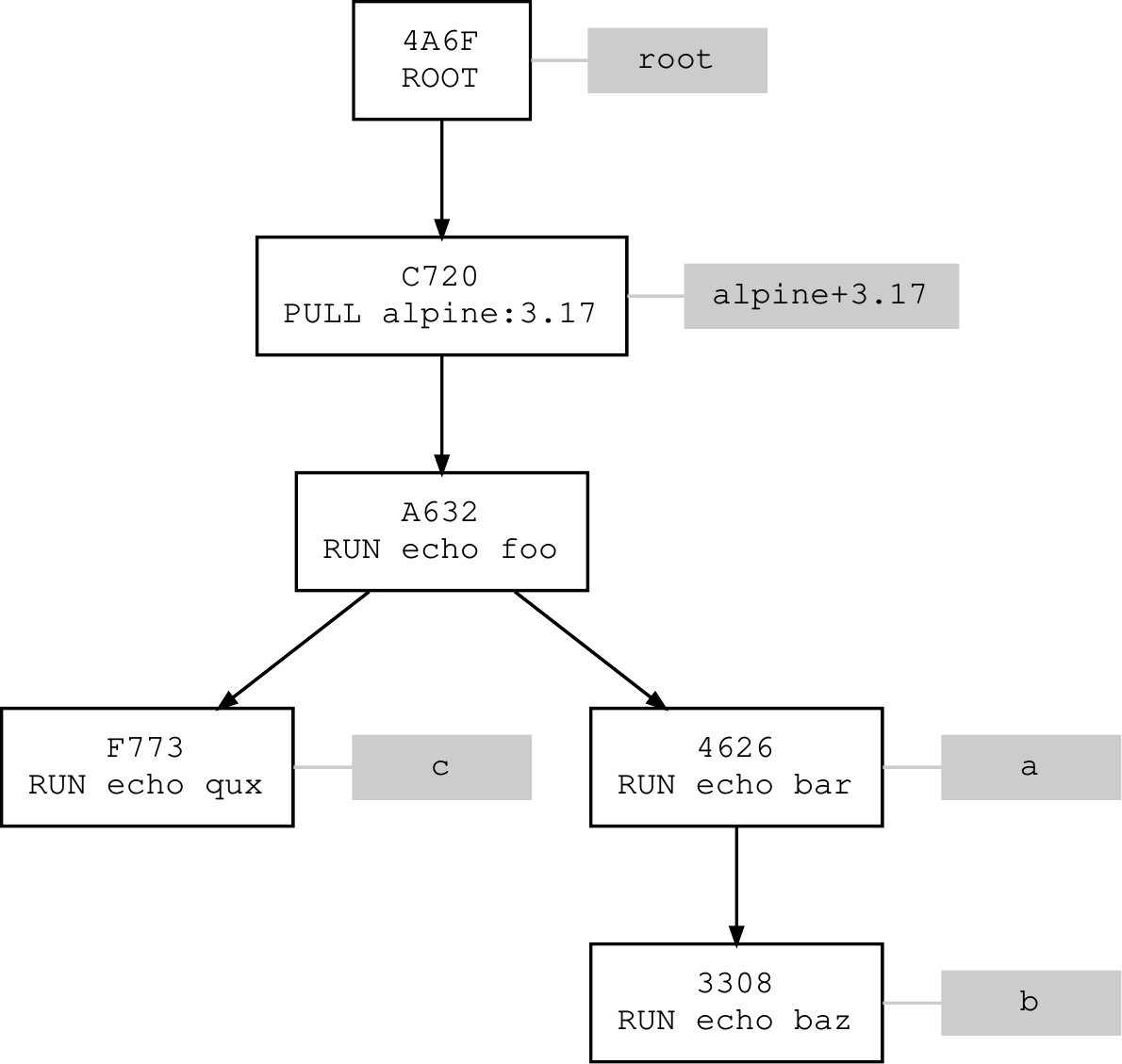}
    };
    \node at (72pt,9pt) {
\begin{dfsmall}{c.df}{16.5ex}
FROM alpine:3.17
RUN echo foo
RUN echo qux
\end{dfsmall}
    };
  \end{tikzpicture}
  \caption{Cache after building (1)~\protect\code{a.df} then \protect\code{b.df} and (2)~\protect\code{c.df}, in either order.}
  \label{fig:buildc}
\end{figure}

Figure~\ref{fig:buildc} shows the cache after also building \code{c}. Assuming \code{a} and \code{b} are already built, then \code{RUN echo foo} is a cache hit and not executed. Next, \code{RUN echo qux} misses, so it is executed and the results stored as \code{F773} at the tip of branch \code{c}. If \code{c} were built before \code{a} and \code{b}, the resulting tree would be the same, but \code{RUN echo foo} would be a cache miss for \code{c}, not \code{a}.

\subsubsection{Build a changed Dockerfile}
\label{sec:change}

\begin{figure}[b]
  \centering
  \begin{tikzpicture}
    \node at (0,0) {
      \includegraphics[scale=\treescale]{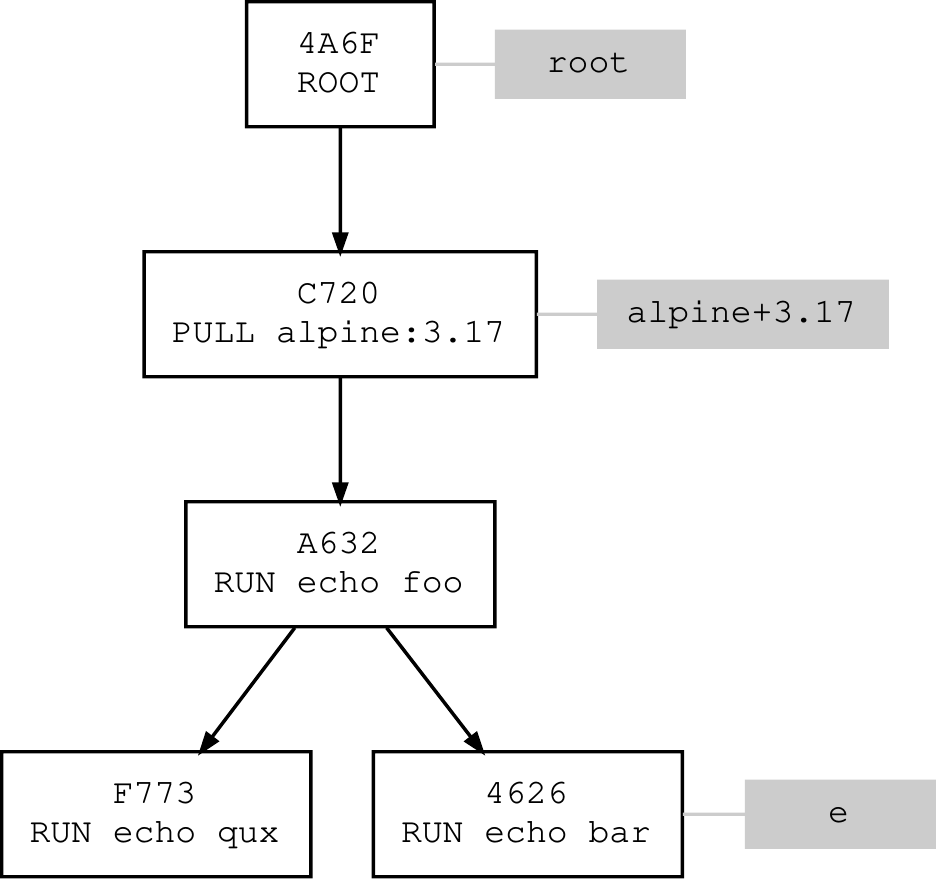}
    };
    \node at (108pt,70pt) {
\begin{dfsmall}{a.df}{16.5ex}
FROM alpine:3.17
RUN echo foo
RUN echo bar
\end{dfsmall}
    };
    \node at (108pt,-18pt) {
\begin{dfsmall}{c.df}{16.5ex}
FROM alpine:3.17
RUN echo foo
RUN echo qux
\end{dfsmall}
    };
  \end{tikzpicture}
  \caption{Cache after building image \protect\code{e} with \protect\code{a.df}, then \protect\code{c.df}, and finally \protect\code{a.df} again.}
  \label{fig:change}
\end{figure}

A common use case while developing Dockerfiles is a change-build loop. Figure~\ref{fig:change} shows (starting from an empty cache) the result of building image \code{e} with \code{a.df}, changing the last instruction of \code{a.df} to create \code{c.df}, using that to build \code{e} again, then reverting the change and building a third time with the original \code{a.df}. (The \id{}s being the same as the other figures is not a coincidence — recall that states are identified by their parent, instruction, and visible input only, which is the same.)

The first build creates the \code{4A6F} → \code{C720} → \code{A632} → \code{4626} path and labels it \code{e}. The second creates the → \code{F773} path (with one cache miss) and moves the label \code{e} to that branch, leaving \code{4626} unlabeled. Finally, the third build finds only cache hits and moves \code{e} back to \code{4626}, leaving \code{F773} unlabeled.\footnote{It is indeed possible to have states with multiple labels. This happens when there are multiple names for the same image.}

\subsubsection{Rebuild, non-unique \id{}s, and search strategy}

\begin{figure}
  \centering
  \begin{tikzpicture}
    \node at (0,0) {
      \includegraphics[scale=\treescale]{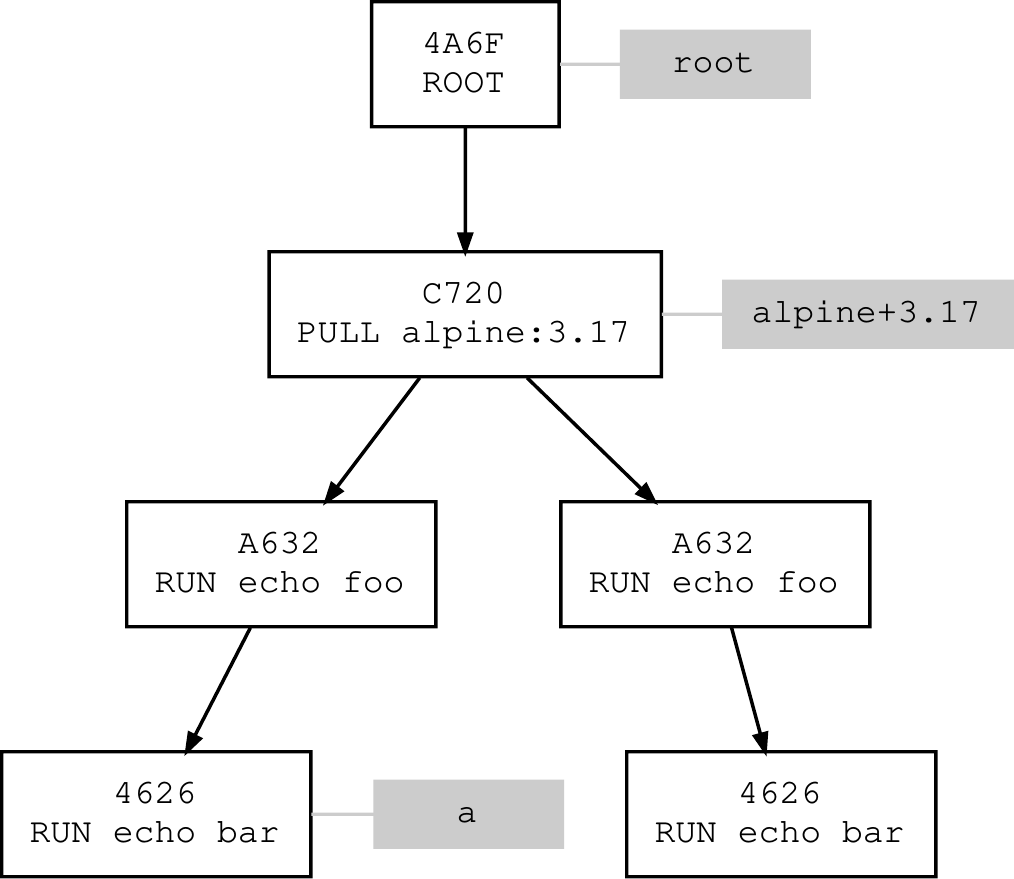}
    };
    \node at (-103pt,60pt) {
\begin{dfsmall}{a.df}{16.5ex}
FROM alpine:3.17
RUN echo foo
RUN echo bar
\end{dfsmall}
    };
  \end{tikzpicture}
  \caption{Cache after building \protect\code{a.df}, then building it again in rebuild mode (\protect\code{--rebuild}). Note the duplicate \id{}s.}
  \label{fig:rebuild}
\end{figure}

Figure~\ref{fig:rebuild} demonstrates an important property of \id{}s: they are not necessarily unique. In this case, the situation comes about because \code{a.df} is first built as image \code{a}, which creates the right-hand (unlabeled in the figure) branch $\texttt{4A6F} \rightarrow ... \rightarrow \texttt{4626}$ and labels it \code{a}. Then, \code{a} is built a second time in rebuild mode (\code{ch-image build --rebuild}), which causes all non-\code{FROM} instructions to be treated as cache misses whether or not they are in the cache. This creates the left-hand branch with the same \id{}s and moves the label \code{a} to it.

Are the two branches really the same? Unlikely, because the user probably had a good reason for rebuild mode, i.e., the steps have \emph{invisible} input (see §\ref{sec:assumptions}) that the user knows about but Charliecloud does not. But, we now have a conundrum because when processing a cache hit, Charliecloud must select a single node from multiple nodes with the same \id{}. Our design goals for this process include (1)~don’t surprise the user, and (2)~do the right thing, i.e., pick the commit the user would expect and want us to pick.

\Id{}s \emph{are} unique within a given branch, and we assume that if the user is building image $x$, they want cache hits most closely related to $x$. The search strategy is therefore: (1)~if rebuilding image $x$, a \id{} match on branch $x$ takes priority, and (2)~otherwise, use the most recently-created matching node anywhere in the cache.

Some alternate search strategies we considered but rejected are:

\begin{itemize}

\item Search only the branch labeled for the current image. This avoids duplicate \id{} problems but can cause false negatives. For example, in Figure~\ref{fig:change}, rebuilding \code{e} with \code{c.df} would not find \code{F773} and \code{RUN echo qux} would be needlessly re-executed.

\item Search the whole cache first, with some global priority such as recency. This is prone to selecting the wrong commit. For example, in Figure~\ref{fig:buildc}, node \code{A632} is shared by all three images. If \code{c} is re-built with \code{--rebuild}, that will create a newer \code{A632}, which would then be used by any rebuilds of \code{a} or \code{b} (in normal mode without \code{--rebuild}). That is, the cache hits of one image could be changed by activity on another image, which seemed too surprising.

\item Search only labeled branches and add more labels to reflect the build history. For example, when rebuilding image \code{foo}, move the branch name \code{foo} to the new branch and rename the old branch to \code{foo_1}. Then, search the named branches in increasing order of age. This seemed too complicated and does not let different-named images share cache entries.

\end{itemize}

\subsection{\Id{} computation in more detail}
\label{sec:stateid}

As discussed above, an image state and its corresponding cache entry, instruction, etc. are identified by a non-unique \vocab{\id}, which is a digest of ancestry and visible input. This computation has significant nuance, so this discussion is a summary and the Charliecloud source code is authoritative.\footnote{In particular, \id{}s may not be stable across Charliecloud versions. The cost is extra cache misses and cache pollution, which seemed an acceptable cost given how hard it would be to maintain stability.}

\Id{}s are 128-bit MD5 digests of the parent \id{} and the instruction’s visible input.\footnote{We use MD5 to emphasize that the cache has not been hardened against malicious alteration. Despite its cryptographic weaknesses, MD5 still has negligible risk of accidental collisions~\cite{rfc6151}.} While Git’s own commit IDs are also digests of input that includes everything we need~\cite{burgdorf2014anatomy}, they are unsuitable for at least two reasons: (1)~Git hashes include a timestamp, which we don’t want, and (2)~\id{}s are not unique but Git hashes are. Digests for different types of states are computed as follows:

\begin{description}

\item[Empty root state.] No parent and pseudo-instruction \code{ROOT}. The \id{} is not a digest of anything but instead simply the constant \code{4A6F:73C3:A9204361:7061626C:616E6361}.

\item[Pulled image.] Parent is the empty state. Pseudo-instruction is \code{PULL <imageref>}, but importantly it is not included in the \id{} computation. Visible input is the \vocab{image manifest} obtained from the image registry. This is a text string that describes the image, including digests computed at push time of the image configuration and all layers, so any change to the actual image in the registry will change the manifest. Therefore, image changes will change the \id{}, but because the the base image name is not included in the digest, the same image with different names will give the same \id{}.

\item[Dockerfile instructions.] This the the normal case. \Id{} is the digest of (1)~parent \id{}, (2)~instruction text, and sometimes (3)~additional input. The instructions with additional input or other non-standard behavior are:

  \begin{description}

  \item[\code{ARG}.] Digested instruction text includes \code{ARG}, the variable name, and the variable value, except for certain variables such as web proxy and SSH agent information, where the value is excluded~\cite[§6.7.4.2]{man-ch-image}. There is no additional input.

    For example, \code{FOO} is a normal variable; given \code{--build-arg=FOO=bar}, the digested instruction text is \code{ARG FOO=bar}. However, the value of \code{HTTP_PROXY} is not digested. The purpose of this exception is to avoid cache misses due to irrelevant environment changes such as the presence of a web proxy (e.g., working on site) or not (e.g., working from home).

  \item[\code{COPY}.] Additional input is basic metadata of all the source files: filename, file type and permissions, size, and last modified time. Unlike Docker, we do not use file contents~\cite[§“Leverage build cache”]{docker2023best}. This has two implications. First, it is possible to change a source file and still hit by manually restoring the last-modified time, but this is hard to do accidentally. Second, \code{COPY} needs much I/O even on miss because it must \code{stat(2)} every source file before checking the cache.\footnote{This could be improved with a two-step scheme, i.e., if the instruction text has changed we know it’s a miss and need not examine the files.} However, this is less I/O than reading the file content too like Docker.

  \item[\code{FROM}.] Recall that \code{FROM} doesn’t change the image; it simply ensures the base image exists and connects it to the current image. Hence, \code{FROM} never appears in the cache.

  \item[Unsupported instructions.] These are ignored by the cache because they have no effect on the image.

  \end{description}

\end{description}

\subsection{Adapting Git for image states}
\label{sec:git-format}

Git is a state-oriented version control system designed to store changing directory tree states as a directed acyclic graph (DAG); many people consider it to be a fancy versioned filesystem~\cite[§1.3]{chacon2023git}. It is mature software that is extremely widely used with both a wide and deep pool of expertise. It is performant, well tested, and supported by well-resourced actors~\cite{stolee2020scalar}.

These properties make Git an appealing base for Charliecloud’s Git cache. (Recall that all trees are are also DAGs.) Further, the Charliecloud team was already familiar with Git because we use it for development, and using Git for its cache fits Charliecloud’s philosophy of using standard tools whenever practical.

The build cache is stored as a bare repository within Charliecloud’s storage directory (subdirectory \code{bucache}). Each image is a Git \vocab{worktree}~\cite{git2022worktree} of this repository (subdirectory \code{img}). Because it is designed for source code rather than arbitrary files\footnote{We use \vocab{file} in the generic sense, i.e., including all file types such as directories, symlinks, etc., unless clear from context or otherwise specified.}~\cite{git2010content}, Git does present several challenges for our use case:

\begin{enumerate}

\item Only regular files, symlinks, and directories are supported. Unprivileged images can also include named pipes (but not sockets or devices), which are ignored by Git.

\item Most file metadata is not supported, including mode, ownership, timestamps, ACLs, and extended attributes (xattrs).

\item Hard links are not supported; each member of a link group becomes an independent regular file with the same content.

\item Empty directories are not supported.

\item Filenames starting with \code{.git} have special treatment.

\end{enumerate}

Another concern is large files. Git uses content-addressable storage~\cite[§10.2]{chacon2023git}, so it must read every stored file in full to compute its digest; that is, files are identified by their contents. For large files, this can be considerable needless work. Instead, if the feature is enabled, Charliecloud identifies files larger than a configurable threshold by their metadata instead: modification time, mode bits, size, and path. Rather than committing these files to Git, Charliecloud stores them out of band (OOB): it moves them out to a separate directory in the work area and commits only the metadata. To restore, it hard-links to this OOB storage, so a given large file can be used by any image.

Charliecloud’s commit and checkout procedures work around these limitations. Specifically, to commit, we:\footnote{Alternatives that we considered but rejected include: (1)~\code{rsync} to a staging directory, which introduces a lot more I/O, and (2)~Git hooks or clean/smudge filters, which lacked features we needed.}

\begin{enumerate}

\item Prepare the image for Git commit:

  \begin{enumerate}

  \item \inheadii{File metadata}: Record.

  \item \inheadii{Hard links to an already-found file}: Record their path and delete. (That is, of a set of hard links pointing to the same inode, we store only the first link encountered and delete the others.)

  \item \inheadii{Large files}: Record their path and move them out of band (to subdirectory \code{bularge}); if already stored out of band, delete.

  \item \inheadii{Empty directories and named pipes}: Delete.

  \item \inheadii{Files starting with \code{.git}}: Rename to \code{.weirdal_}.\footnote{We place Git information at \code{/ch/git} within the image, not the default \code{/.git}, so this does not disrupt Charliecloud’s Git information.}

  \end{enumerate}

\item Save the recorded metadata in a Python pickle file within the image at \code{/ch/git.pickle}.

\item Commit all changes in Git.

\item Restore filenames and deleted files so the image is ready for the next instruction.\footnote{Corollary: worktrees are always dirty from Git’s perspective except during Charliecloud’s commit process.}

\end{enumerate}

We run this procedure for instructions that alter image data, e.g.\ \code{RUN}. For metadata instructions (e.g.\ \code{WORKDIR}), we simply commit the altered metadata. Checkout is simply the reverse, restoring the attributes that were recorded.

Finally, Charliecloud takes advantage of Git’s de-duplication capabilities. Identical files are de-duplicated at commit time, in-band (small) files by Git based on content and OOB (large) files by Charliecloud based on metadata. Similar in-band files are de-duplicated upon cache \vocab{compaction}, which invokes Git’s garbage collection. Git compares file content and stores similar files as base data plus deltas (i.e., diffs).\footnote{Git sorts files by metadata and compares only within a sliding window, avoiding $O(n^2)$ behavior~\cite{git2022repack}.} Similar OOB files are not de-duplicated (because Git never sees their content). Compaction also deletes image states and OOB files no longer used by any named image.

We next turn to an evaluation of this cache’s performance.

\section{Performance}
\label{sec:performance}

Charliecloud’s build cache works, but that’s not enough — performance also must be acceptable. The cache must be (1) fast enough and (2) efficient enough with disk space. This section details experiments showing that the cache not only has acceptable performance but is superior to layered caches in some circumstances. We also tested performance by filesystem and large-file threshold. Analysis code and raw data are in the supplemental information.

\subsection{Experiment procedure}

\subsubsection{Overview}

% https://memory.net/product/m393a2k40cb2-cvf-samsung-1x-16gb-ddr4-2933-rdimm-pc4-23466u-r-single-rank-x4-module/
% https://www.newegg.com/samsung-pm1725b-1-6tb/p/2U3-0005-000H3

Our test cluster had one head node and 8 compute nodes running AlmaLinux~8.7, each with (1)~one 32-core AMD EPYC 7502 CPU, (2)~\qty{128}{GiB} of memory in \num{8} DDR4 DIMMs, specified as \qty{31}{ns} latency, \qty{2933}{MT/s}, \qty{47}{GiB/s} bandwidth, (3)~one \qty{1.5}{TiB} Samsung NVMe disk specified at \qty{800}{kIOPS} read, \qty{190}{kIOPS} write and \qty{3.4}{GiB/s} sequential read, \qty{3.0}{GiB/s} write. The interconnect was Mellanox ConnectX-5 InfiniBand running at \qty{100}{Gb/s}. We used Git v2.40.0 compiled from source with “profile feedback” optimizations.

We tried to avoid builds downloading anything from the internet, in order to isolate the experiment from such variability. On the head node, we ran a local image repository (Docker Hub \code{registry:2} image\footnote{\url{https://hub.docker.com/_/registry}}) to hold base images, and we ran a Squid caching proxy that handled all web traffic from the compute nodes. However, given DNF’s use of many different mirrors, we suspect this isolation was not entirely effective, leading to noisier results for RPM-based images. (Turning off mirroring led to random build failures.)

The experiment had five factors: (1)~container implementation used for building, (2)~filesystem and backing storage of the implementation’s work area, (3)~image being built, (4)~large-file threshold (Charliecloud only), and (5)~\vocab{cache temperature}, i.e., what was already in the cache before the build. We wrote a custom test driver in Python to iterate through these factors, running 8 tests in parallel (one per node) for local filesystems and 4 for NFS (the other 4 nodes serving the filesystem), with a total of 24 repetitions per condition. Disk caches were cleared before each test. Source code for the driver is available in the supplemental data. The following sections detail these factors.

\subsubsection{Container implementation}

We report results for four implementations, listed here by the abbreviations used in this paper:

\begin{enumerate}

\item \inhead{ch}: Charliecloud 0.33, which is our own container implementation described above; for more detail, see~\cite{priedhorsky2021privilege, priedhorsky2017sc}. It is a fully unprivileged (Type III~\cite{priedhorsky2021privilege}), HPC-focused implementation of approximately \qty{20}{k} lines of code (LOC), as measured with \cite{jolav-countloc}. Charliecloud uses the novel Git-based build cache detailed above.

\item \inhead{ch–}: Charliecloud 0.33 with build cache disabled

\item \inhead{dko}: Docker 23.0.2 with \code{overlay2} storage driver, which uses Linux’ in-kernel OverlayFS implementation and is “the preferred storage driver” in all cases we can discern~\cite{docker2023storage}. Docker is a general-purpose container implementation with a client/daemon architecture, which confuses HPC scheduling tools because containers are children of the daemon, not the \code{docker run} command. Docker is approximately \qty{470}{kLOC},\footnote{\label{fn:novendor}LOC for Docker and Podman excludes subdirectory \code{vendor}, which in Golang projects is third-party code.} about 23 times larger than Charliecloud. We ran it in the default, privileged (Type~I) mode.

  % (Docker does have a “rootless” partly-unprivileged Type~II mode that appears to us little-used.)

\item \inhead{pmo}: Podman 4.4.3 with its analogous \code{overlay2} storage driver. This general-purpose implementation is designed to duplicate Docker’s command-line interface but without a daemon~\cite{henry2019podman}, producing an HPC-friendly process tree. Podman is approximately \qty{210}{kLOC},\textsuperscript{\ref{fn:novendor}} $10{}\times$ Charliecloud. We ran it the default Type~I mode.

\end{enumerate}

We did pilot-test Docker and Podman with their FUSE-OverlayFS storage drivers, which use OverlayFS implemented in user-space via FUSE. However, we do not report these results, which were generally slower than kernel OverlayFS. Kernel OverlayFS is the best practice and starting in Linux~5.11 (February 2021) is available to unprivileged processes~\cite{calleja2021linux511}. This kernel is already making its way into HPC-relevant distributions; e.g., SUSE Enterprise Linux 15~SP4 (June 2022) has kernel 5.14~\cite{moutoussamy2022sle15sp4}. In short, we felt that comparing Charliecloud to a technology that performs worse and may soon be obsolete was not a good use of column inches.

\subsubsection{Filesystem}

We tested container build with three work area storage configurations:

\begin{enumerate}

\item \inhead{ext4}: Backed by the NVMe drive, size \qty{1.5}{TiB}.

\item \inhead{tmpfs}: Memory-backed, size \qty{63}{GiB}. \qty{2.3}{GiB} were used by the host operating system, leaving \qty{61}{GiB} for work areas.

\item \inhead{NFS}: Each pair of nodes exported their ext4 filesystem to each other, using NFSv3 over IP over IB.

\end{enumerate}

\subsubsection{Test image}

\begin{table}
  \caption{Test images built by our experiment. \vocab{Ins} is the number of instructions and \vocab{MiB} is final image size in megabytes.}
  \begin{tabular}{@{}lrrl@{}}
    \toprule
      \multicolumn{1}{c}{\textbf{name}}
    & \multicolumn{1}{c}{\textbf{ins.}}
    & \multicolumn{1}{c}{\textbf{MiB}}
    & \multicolumn{1}{c}{\textbf{description}}
    \\
    \midrule
    \multicolumn{4}{@{}l@{}}{\textit{images from Charliecloud test suite}} \\
    almalinux &   4 &   560 & basic image for compiling programs \\
    openmpi   &  17 &   740 & common HPC library \\
    paraview  &  23 & 1,900 & long-ish application build \\
    \midrule
    \multicolumn{4}{@{}l@{}}{\textit{small synthetic images}} \\
    micro     &   2 &     7 & minimal with small base (Alpine) \\
    mini      &   2 &   200 & minimal with large base (AlmaLinux) \\
    \midrule
    \multicolumn{4}{@{}l@{}}{\textit{large synthetic images}} \\
    megainst  & 129 &     7 & many instructions \\
    megafiles &   6 & 4,200 & many small files ($2^{18} \times \qty{16}{KiB}$)\\
    megabytes &   6 & 4,200 & few large files ($32 \times \qty{128}{MiB}$) \\
    megapkg   &   9 & 8,200 & many distro packages (2,561 of them) \\
    \bottomrule
  \end{tabular}
  \label{tab:images}
\end{table}

Table~\ref{tab:images} summarizes the nine images we built. The actual Dockerfiles are in the supplemental data, and \code{almalinux} is also shown in Figure~\ref{fig:dockerfile}. The images range in size from \qty{7.0}{MiB} to \qty{8.2}{GiB} and the Dockerfiles from 2 to 129 instructions. Three images (\code{almalinux}, \code{openmpi}, and \code{paraview}) are from the Charliecloud test suite.\footnote{\code{paraview} is a superset of \code{openmpi}, which in turn is a superset of \code{almalinux}. In the Charliecloud source code these images build on each other, but for this experiment we copied the instructions so every Dockerfile started from an external base image.} The other six are synthetic, written for this paper to exercise the build cache in specific ways.

\subsubsection{Large-file threshold (Charliecloud only)}

Recall that Charliecloud can store large-files out of band, i.e., outside the Git repository. We tested seven largeness thresholds: from \qty{1}{MiB} (i.e., files larger than \qty{1}{MiB} are stored OOB) to \qty{32}{MiB} by powers of 2, as well as disabled (i.e., no files stored out of band).

\subsubsection{Cache temperature}

This factor defines how much of the image has already been built and cached: none (\vocab{cold}), all (\vocab{hot}), or about half of it (\vocab{warm}). While the previous factors are independent, temperature is not. The innermost experiment loop is:

\begin{enumerate}

\item Delete the work area, if one exists.

\item Initialize a new work area. For Charliecloud we did \code{ch-image list}; for Docker and Podman we built a trivial image (\code{FROM scratch}).

\item Build the test image. (\vocab{Cold cache}.)

\item Build it again. (\vocab{Hot cache}.)

\item Build it again, but with an instruction roughly halfway through (selected manually by us) modified to be functionally identical but look different to the cache. Specifically, we change the trailing comment \code{#WARM#} to \code{&& true}; see line~7 of Figure~\ref{fig:dockerfile}. (\vocab{Warm cache}.)

\end{enumerate}

The next four subsections detail our results.

\subsection{Build time per implementation}
\label{sec:build-time}

\subsubsection{Introduction}

This section asks which implementation builds the fastest. For Charliecloud, we use a large-file threshold of \qty{4}{MiB} because it seemed a reasonable moderate choice (see §\ref{sec:large-file} below).

Notably, Git-based cache builds can have a \vocab{cooldown time} while Git garbage-collects in the background. The only ext4 builds where this time was more than a fraction of a second was cold-cache \image{megafiles} at \qty{30}{s} and \image{megapkg} at \qty{80}{s}. Background garbage collection can also be killed without ill effect. Thus, cooldown seemed to us a minor effect and we do not analyze it here, but full data are in the supplemental information.

\subsubsection{Cold cache}

\begin{figure}
  \centering
  \begin{tabular}{@{}r@{}}
            \includegraphics{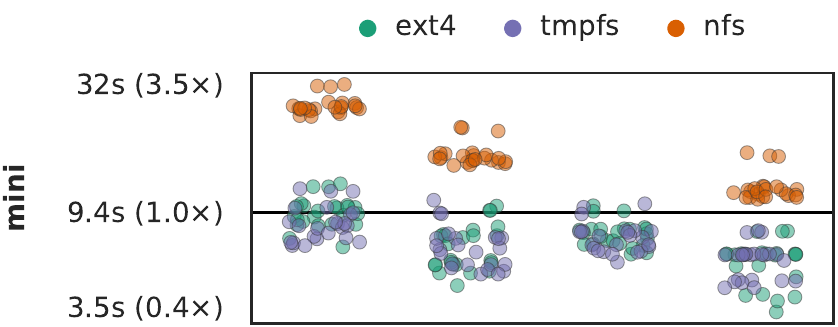}
    \\[3pt] \includegraphics{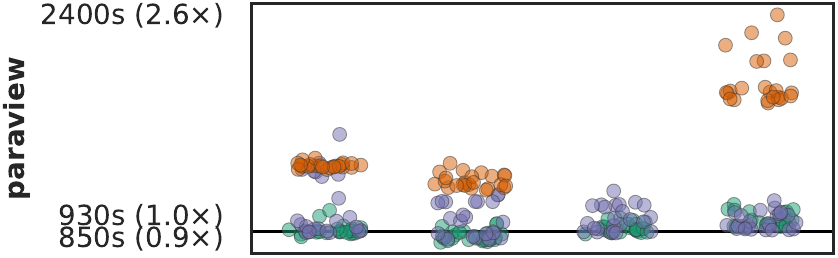}
    \\[3pt] \includegraphics{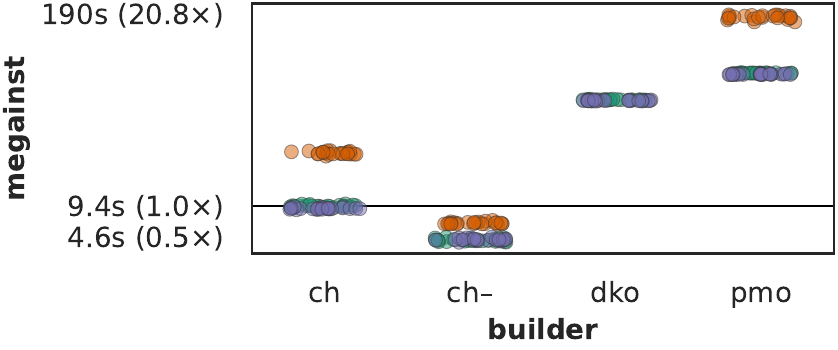}
  \end{tabular}
  \caption{Cold build times of three selected images for Charliecloud (\ch), Charliecloud with cache disabled (\chd), Docker using overlayfs (\dko), and Podman using overlayfs (\pmo). X~axis is the four implementations tested; log-scale Y~axis is time, both in seconds and relative to median Charliecloud on ext4 (lower is better). Each dot represents one test.}
  \label{fig:build-eg}
\end{figure}

Figure~\ref{fig:build-eg} shows detailed times for cold-cache builds of three representative images. For image \image{mini}, Charliecloud is slower than Docker (here by median 1.2×) and Podman (1.5×).\footnote{We suspect the noisiness of this image is caused by response time of our container registry, as it is serving up to 8 requests in parallel and just a few seconds’ delay is a non-trivial fraction of this image’s build.} For \image{paraview}, the three implementations are roughly equal, with the medians all within 4\% of one another. For \image{megainst}, Charliecloud is considerably faster than Docker (5×) or Podman (7×).

\begin{table*}
  \caption{Median build time on ext4 for all images and all cache temperatures. Column \vocab{time} lists build time in seconds, while \vocab{vs.\ ch} lists the percentage difference from Charliecloud, both to two significant figures. Blues indicate faster than Charliecloud (clipped at –75\%, i.e.\ 4× faster), reds slower (clipped +300\%, i.e.\ 4× slower), i.e., redder is better from Charliecloud’s perspective.}
  \input{figures/smmy_build.tex}
  \label{tab:smmy-build}
\end{table*}

Table~\ref{tab:smmy-build} then summarizes build time for all images and cache temperatures on ext4. Charliecloud with cache disabled (which we treat as cold) is consistently the fastest. This expected as there is no cache overhead at all; the build decays to simply running each instruction one after another in the work area.

Charliecloud’s cache-enabled build time vs.\ the other two implementations falls into three rough regimes. First, \image{megafiles} and \image{mini} are somewhat slower, up to 1.7× (–41\%). Recall that \image{megafiles} is \qty{4}{GiB} of small files, which all must be read in full by Git in order to compute their identifying digests; \image{mini}’s build time is dominated by downloading, unpacking, and committing to Git a \qty{190}{MiB} (uncompressed) tarball. Neither is favorable conditions for Git.

Second, a group of five images — \image{almalinux}, \image{megapkg}, \image{openmpi}, \image{paraview}, and \image{micro} — shows similar performance for all three implementations, Charliecloud ranging from 4.4\% slower to 23\% faster. Notably, all four images emphasizing installation of real software are in this group: \image{almalinux} installs 361 RPM packages, \image{megapkg} installs 2,600 Debian packages, and \image{openmpi} and \image{paraview} spend most of their time building software from source.

The final group is images where Charliecloud builds faster than the competition. For \image{megabytes}, Charliecloud’s \qty{21}{s} is 1.5× faster than Docker’s \qty{31}{s} and 3× faster than Podman’s \qty{70}{s}. This highlights the value of Charliecloud’s OOB large file storage: the bulk of the image is 32 \qty[round-mode=none]{128}{MiB} random files, which exceed the \qty{4}{MiB} large-file threshold and require only a few metadata system calls to store. Without this optimization (i.e., infinite large-file threshold), Charliecloud requires median \qty{47}{s} to build the image.

Charliecloud’s best performance, \image{megainst}, is \code{FROM} a single-layer \code{alpine:3.16} followed by 128 \code{RUN echo}, i.e., the image is 129 instructions deep. This poses quite a challenge for overlay-based caches, but Charliecloud’s flat build is agnostic to instruction count. (Charliecloud’s \qty{9.4}{s} median build time is 5× faster than Docker’s \qty{46}{s} and 7× faster than Podman’s \qty{68}{s}.) This improves usability — users can organize their Dockerfiles in the way that is clearest and best fits the project, without concern about minimizing layers.

\subsubsection{Warm cache}

This is a similar story to the cold cache. The Charliecloud-slower and Charliecloud-faster regimes have grown slightly, and the baselines are faster. Note that not all the images have a instruction near the halfway point available for the warm modification; for example, Charliecloud builds \image{almalinux} in \qty{4.8}{s} warm, less than 10\% of the \qty{54}{s} cold time, as opposed to the better balanced \image{megapkg} (\qty{300}{s} is 60\% of \qty{490}{s}).

\subsubsection{Hot cache}

On the other hand, Charliecloud’s relative performance for hot cache is consistently quite poor (with one outlier), up to 50× slower than Docker (e.g.\ on \image{megafiles}). We suspect this is due to inefficiencies in the code that finds cache hits and are working to optimize these operations. Fortunately, the bad performance is relative to a fast baseline. While waiting a few to several seconds for a no-op build is not ideal, we believe it is acceptable.

\subsubsection{Verdict: Fast enough, sometimes faster}

Charliecloud’s Git-based build cache appears to be competitive with Docker and Podman’s overlay-based cache when cold and warm, with similar build times on some images, somewhat slower on others, and considerably faster in regimes that may enhance usability. Charliecloud’s build time with hot cache is consistently much slower, but the baseline is faster and there may be opportunities for optimization. We also observe that Podman is generally slower than Docker, which is not good news for HPC as Podman’s process structure is more favorable for scheduling.

\subsection{Storage space per implementation}

\subsubsection{Introduction}

This question compares the disk usage of Charliecloud to other implementations.
Here, Charliecloud’s results also use the \qty{4}{MiB} threshold ext4 only because disk usage does not vary meaningfully by filesystem. We also omit a hot cache discussion because it is essentially identical to cold. The final omission is Podman, which performed very similarly to Docker. (Full data are in the supplemental information.)

For this analysis, it is helpful to understand the structure of Charliecloud’s work area. Nearly all disk use is in four subdirectories:

\begin{enumerate}

\item \inhead{\code{img}}: Uncompressed, unpacked copy of each named image’s current state.

\item \inhead{\code{dlcache}}: Manifests and blobs downloaded from container registries, stored verbatim as received.

\item \inhead{\code{bucache}}: Build cache Git repository.

\item \inhead{\code{bularge}}: Build cache out-of-band large files.

\end{enumerate}

\begin{table*}
  \caption{Storage usage after each build for several flavors of Charliecloud compared to Docker. As in Table~\ref{tab:smmy-build}, bluer indicates better (smaller) than the baseline and redder indicates worse (larger), but here the baseline is Docker (\dko), i.e.\ Charliecloud wants bluer. Columns \vocab{MiB} are usage in megabytes, while \vocab{vs.\ dko} is a percentage comparison to Docker; \vocab{time} is compaction time. See text for discussion of the other column labels.}
  \input{figures/smmy_disk.tex}
  \label{tab:smmy-disk}
\end{table*}

Table~\ref{tab:smmy-disk} summarizes Charliecloud’s storage usage; unlike §\ref{sec:build-time}, the baseline is Docker, so favorable for Charliecloud are negative percentages (i.e., bluer). We show three flavors of Charliecloud:

\begin{enumerate}

\item \inhead{\chd} is cache-disabled, i.e., \code{img} and \code{dlcache} only.

\item \inhead{\ch} is cache-enabled, directly after build (and cooldown).

\item \inhead{\chgc} is cache-enabled with a cache compaction cycle after build.

\end{enumerate}

The unpacked images in \code{img} present a clear opportunity for storage tuning. Therefore, we present two usage versions: \vocab{everything} is the full storage directory, while \vocab{w/o unpacked} excludes the unpacked working images (i.e., \code{img}), leaving just the layer tarballs and build cache.

\subsubsection{Cold cache}

We first consider cache disabled (\chd), to explore how disk usage of the basic Charliecloud build process compares to Docker. Unlike build time, where \chd{} is consistently faster, here it ranges from about 2½× as large as Docker (\image{mini} at +140\%) to roughly the same (e.g.\ \image{megapkg}). The reason for this large penalty on \image{mini} is that Charliecloud stores three copies of essentially the same image: the unpacked base image (\image{almalinux:8.7}), the unpacked final image, and the compressed image layer(s) for the base image downloaded from the image registry. This is also true for \image{megapkg}, but that final image is much larger than the base, so it dominates. Docker, on the other hand, based on examination of its work area, appears to store approximately one uncompressed copy, though fragmented into many subdirectories.

With cache enabled (\ch), the full work area (column \vocab{everything}) adds a fourth image copy, the compressed one in Git. \image{mini} goes to +200\% vs.\ Docker and \image{megapkg} +23\%. Excluding unpacked images (\vocab{w/o unpacked}) yields \image{mini} of just +34\%, and \image{megapkg} is now roughly half the size of Docker, –40\%. We suspect this is due to Git’s object compression and/or file de-duplication (recall that Git stores identical files only once).

Finally, Charliecloud’s cache compaction (\vocab{ch compacted}) sometimes has benefit even here, with only a single built image in the cache. For example, we can spend \qty{4.0}{s} to reduce \image{mini} from +200\% to +160\%; without the unpacked images, the benefit is greater, +34\% to –13\%. The best relative improvement is \image{megainst}, \qty{1.0}{s} for +120\% down to +43\% (or +22\% to –52\% excluding unpacked). Some images, however, show no benefit; for example, we can spend \qty{82}{s} compacting \image{megafiles} and get nothing. This particular case is unsurprising because it is specifically designed to thwart Git, but \image{megapkg} and \image{megabytes} also do not shrink.

\subsubsection{Warm cache}

% NOTE: This figure belongs quite a bit later but isn’t being placed correctly.
\begin{figure*}
  \centering
  \begin{tabular}{@{}ccc@{}}
      \includegraphics{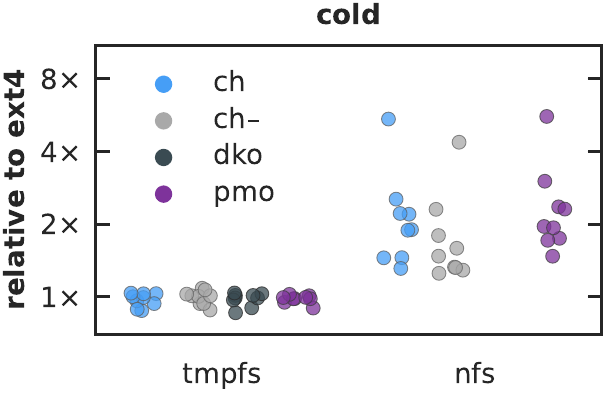}
    & \includegraphics{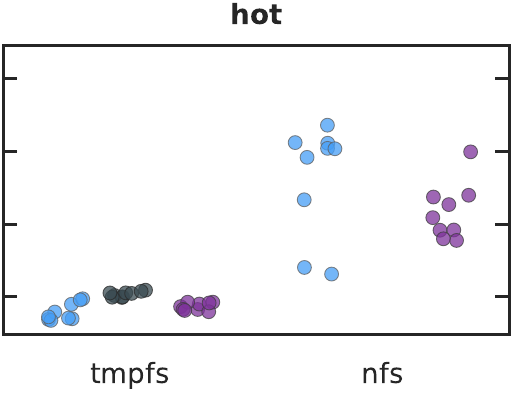}
    & \includegraphics{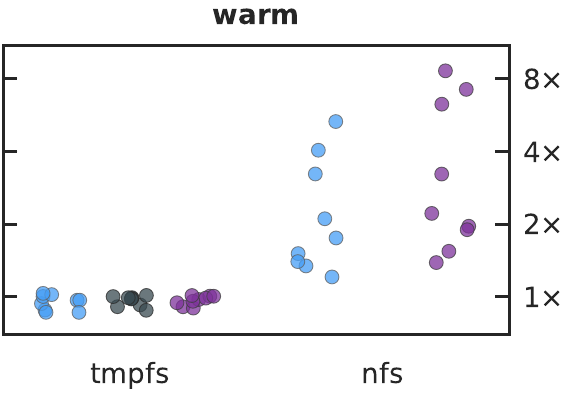}
  \end{tabular}
  \caption{Build time relative to ext4. Each dot is the median of an image’s 24 repetitions. tmpfs was modestly slower than ext4, while NFS was quite a bit slower as well as noisier.}
  \label{fig:filesystems}
\end{figure*}

Recall that the warm condition makes a no-op modification of an instruction and then builds the new Dockerfile. This creates two images that the build cache thinks are different (because of the different instruction sequence to create them) but really are the same (because the instruction sequences are semantically identical). The image from the cold-cache step is now unnamed. Normally, Charliecloud’s cache garbage collection would delete this unnamed image, but we manually added a Git tag to prevent this, reasoning that real use would not garbage-collect as aggressively and we didn’t want to give Charliecloud an unfair advantage. The warm build does happen after the post-cold-build compaction, so these \ch{} results do enjoy that space savings.

Results for Charliecloud are generally improved even before a second round of garbage collection. The full work area (\vocab{everything}) now ranges from +160\% to –23\% vs.\ Docker, and excluding unpacked is now consistently the same or smaller, up to –62\%.

One image is notable, \image{megapkg}. The Docker work area grew from \qty{7.8}{GiB} cold to \qty{13}{GiB} warm despite essentially identical content of the old and new images. On the other hand, Charliecloud’s grew quite modestly, from \qty{9.6}{GiB} to \qty{9.8}{GiB} (–23\% vs. Docker), or without unpacked \qty{4.7}{GiB} to \qty{4.9}{GiB} (–62\% vs. Docker). We suspect this is the result of Git’s better de-duplication capability.

Compaction offered limited value here. Most images had little to no benefit, and the one that did (\image{megainst}) was quite small already.

\subsubsection{Verdict: Only sometimes competitive so far, but promising}

In these specific experiments, considering its full work area, Charliecloud is indeed usually larger. However, compaction does seem effective. For example, when a second image is introduced in the warm-cache condition, Charliecloud’s relative situation improves, and when excluding unpacked images, Charliecloud’s warm work area is consistently the same or smaller than Docker.

We suspect the poor performances are due to redundancy in Charliecloud’s present default configuration. That is, building a single image causes up to four images to be added to the work area: (1)~unpacked base image, (2)~unpacked final image, (3)~base image layers downloaded from image registry, and (4)~image states from base to final in build cache. This situation arose organically as Charliecloud matures rather than being designed. Now may be the time to analyze the situation more carefully, to understand the trade-offs in play, and modify Charliecloud accordingly.

On the other hand, we suspect the better performances are due to Charliecloud’s broader de-duplication across the entire cache, rather than narrowly between states with an ancestry relationship, as in overlay-based caches. Given that savings often only appeared after the second image was built, we hypothesize the redundancy problem will decrease and the de-duplication benefit will increase in caches with more complex use patterns. That is, it seems quite plausible the Git-based build cache is usually smaller than the overlay-based cache in real use.

\subsection{Build time per filesystem}
\label{sec:filesystem}

Figure~\ref{fig:filesystems} shows build time of tmpfs and NFS relative to ext4. tmpfs showed little advantage over ext4, at most 20\% speedup despite specified performance many times greater. On the other hand, NFS varied from similar performance (20\% slower than ext4) to much worse (9× slower). Docker with OverlayFS does not support NFS, so it has no results here.\footnote{Interestingly, Docker with FUSE-OverlayFS \emph{does} support NFS, while Podman is the opposite: OverlayFS supports NFS and FUSE-OverlayFS does not.} None of the builders showed a storage-related advantage over the others.

\subsection{Effects of large file threshold}
\label{sec:large-file}

This final section analyzes the value of Charliecloud’s out-of-band large-file storage, where files larger than some threshold (user-configurable on a per-build basis) are optionally stored outside the Git cache and hard-linked into unpacked images.

\begin{figure}
  \centering
  \includegraphics{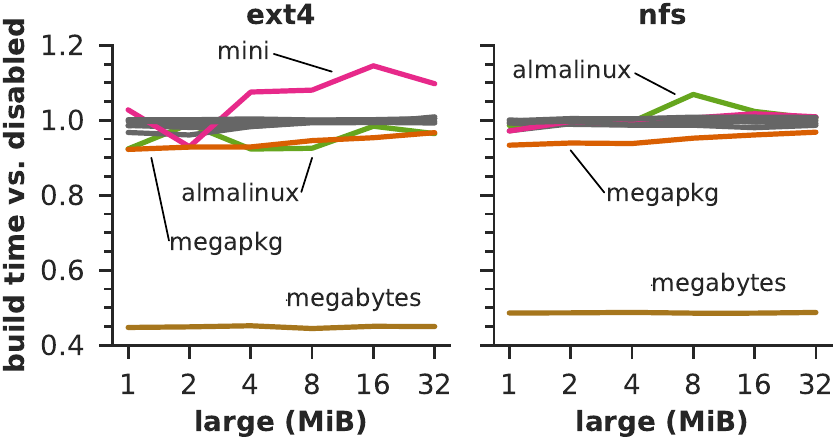}
  \caption{Cold-cache build time of various large-file thresholds, relative to out-of-band large files disabled. This was usually a modest effect.}
  \label{fig:large-build}
\end{figure}

Figure~\ref{fig:large-build} shows the cold build time of large-file thresholds from \num{1} to \qty{32}{MiB} relative to the feature being disabled (warm cache had essentially the same results and hot no effect). The effect is usually minimal, regardless of filesystem. A few highlighted images do show a modest speedup (at most roughly 10\%) that seems to increase toward lower thresholds (i.e., more files stored outside Git). An exception is \image{megabytes}, consistently just over twice as fast because nearly all data are in files larger than all thresholds. This result does show that the feature has clear build speed benefit for images with a notable fraction of data in large files.

\begin{figure}
  \centering
  \includegraphics{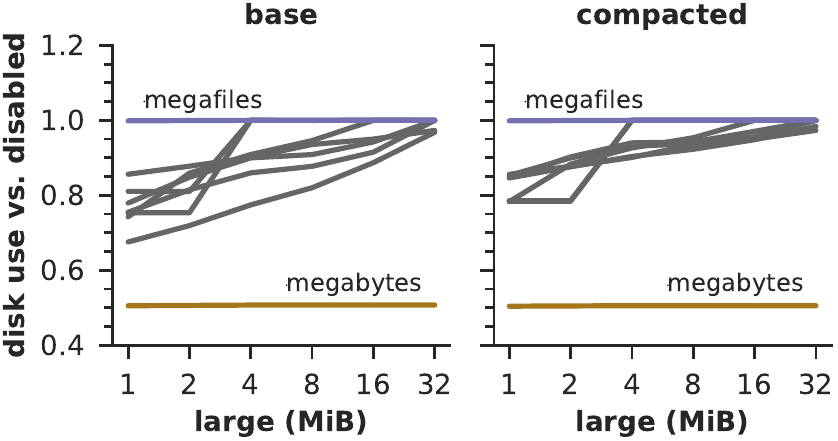}
  \caption{Cold-cache disk usage of various large-file thresholds, before and after garbage collection, relative to OOB large files disabled. This showed greater benefits for most images on lower thresholds.}
  \label{fig:large-disk}
\end{figure}

Figure~\ref{fig:large-disk} shows relative disk usage for the different thresholds, again cold, before and after compaction. Here, there are size benefits for for most images, maximally at the lowest threshold tested and largely disappearing by \qty{32}{MiB}. This is due to large files being stored only once, with two hard links (one in OOB storage and the other in the unpacked image). Compaction reduces the relative benefit because it shrinks the Git cache (the denominator). There are now two exceptions. \image{megabytes} is almost exactly half size and for the same reason: most of its data is always OOB. \image{megafiles} has no benefit, and for the opposite reason: almost all of its data are in files below the smallest threshold and thus never OOB.

While these disk usage results show a clear benefit for a \qty{1}{MiB} (or smaller!) large-file threshold, we used a middle-of-the-road \qty{4}{MiB} for reporting Charliecloud results in previous sections. This is because are concerned they may not generalize. OOB large-files prevent Git’s own de-duplication (of similar files and identical files with different metadata). We suspect that in real caches this will become more important; more experiments will help clarify.

\section{Implications}

We present a new approach for container build caching, based on Git rather than layered filesystems. We explored this approach in the context of a working open-source implementation in Charliecloud, available since version 0.28 in June 2022.

Based on these first-look experiments, the Git-based cache is competitive. On build time, the two approaches are broadly similar, with one or the other being faster depending on context. Both had performance problems on NFS. Notably, however, the Git-based cache was much faster in our many-instruction condition. On disk usage, the winner depended on the condition. For example, we saw the layered cache storing large sibling layers redundantly; on the other hand, the Git-based cache has some obvious redundancies as well, and one must compact it for full de-duplication benefit. However, Git’s de-duplication was effective in some conditions and may prove even better in more realistic situations.

From a structural perspective, the Git-based cache offers potentially significant benefits in three key areas: (1)~representation of container image diffs, which is Git’s \emph{raison d’être} but an awkward fit for tarball layers; (2)~cache overhead, which for Git is imposed only at instruction commit time, is roughly proportional to commit size, and independent of instruction count, while layered caches have overhead proportional to instruction count on every file metadata operation; and (3)~de-duplication, which for Git is done across the entire image cache, with identical files de-duplicated at commit time (for file content if stored in Git or metadata if stored OOB) and similar files at compaction time (which does consume resources and isn’t done for OOB files), while layered caches de-duplicate only identical files on layers with ancestor/descendant relationship.

That is, we believe these results show that the Git-based build cache is highly competitive with the layered approach, with no obvious inferiority so far and hints that it may be superior on important dimensions. We plan to explore it further in Charliecloud, and we hope to support other teams doing so with other layer-free container image builders.

\begin{acks}
  \sloppy

  Vanessa Sochat gave detailed feedback on an early version of this paper, which led to dramatic improvements. This work was supported in part by the \grantsponsor{ecp}{Exascale Computing Project}{} (\grantnum{ecp}{17-SC-20-SC}), a collaborative effort of the U.S.\ Department of Energy (DOE) Office of Science and the National Nuclear Security Administration (NNSA); the Advanced Simulation and Computing Program (ASC); and the LANL Institutional Computing Program, which is supported by the U.S.\ DOE’s \grantsponsor{nnsa}{NNSA}{} under contract \grantnum{nnsa}{89233218CNA000001}. LA-UR~23-29388.

\end{acks}

\sloppy
\printbibliography

\end{document}

%% file: figures/smmy_build.tex
\begin{tabular}{llrrrrrrr}
\toprule
&
& \multicolumn{1}{c}{\textbf{ch}}
& \multicolumn{2}{c}{\textbf{ch–}}
& \multicolumn{2}{c}{\textbf{dko}}
& \multicolumn{2}{c}{\textbf{pmo}}
\\
\textbf{temp} & \textbf{image}
& time
& time
& vs.\ ch
& time
& vs.\ ch
& time
& vs.\ ch
\\

\cmidrule(r){1-2}
\cmidrule{3-9}

\multirow{9}{*}{\textbf{cold}}
& megafiles
&  \qty[round-mode=figures,round-precision=2]{76.10148537158966}{\second}
& \cellcolor[rgb]{0.42139900943076225,0.6768844562046273,0.8189293710563811} \qty[round-mode=figures,round-precision=2]{29.15944230556488}{\second}
& \cellcolor[rgb]{0.42139900943076225,0.6768844562046273,0.8189293710563811} \qty[print-implicit-plus,round-mode=figures,round-precision=2]{-61.683478}{\percent} % 0.383165
& \cellcolor[rgb]{0.7274924425748776,0.8511130711415673,0.914450701464746} \qty[round-mode=figures,round-precision=2]{44.92697322368622}{\second}
& \cellcolor[rgb]{0.7274924425748776,0.8511130711415673,0.914450701464746} \qty[print-implicit-plus,round-mode=figures,round-precision=2]{-40.964394}{\percent} % 0.590356
& \cellcolor[rgb]{0.9833457719881493,0.8981749101024492,0.8480674853563109} \qty[round-mode=figures,round-precision=2]{97.4240951538086}{\second}
& \cellcolor[rgb]{0.9833457719881493,0.8981749101024492,0.8480674853563109} \qty[print-implicit-plus,round-mode=figures,round-precision=2]{28.018651}{\percent} % 1.280187
\\
\bottomrule
\end{tabular}

%% file: figures/smmy_disk.tex
\begin{tabular}{llrrrrrrrrrrrr}

      \toprule
      &
      & \multicolumn{1}{c}{\multirow{2}{*}{\textbf{dko}}}
      %& \multicolumn{2}{c}{\multirow{2}{*}{\textbf{pmo}}}
      & \multicolumn{2}{c}{\textbf{ch–}}
      & \multicolumn{4}{c}{\textbf{ch}}
      & \multicolumn{5}{c}{\textbf{ch compacted}}
      \\
      &&
      %&&
      & \multicolumn{2}{c}{everything}
      & \multicolumn{2}{c}{everything}
      & \multicolumn{2}{c}{w/o unpacked}
      &
      & \multicolumn{2}{c}{everything}
      & \multicolumn{2}{c}{w/o unpacked}
      \\
      
\textbf{temp} & \textbf{image}
& MiB
& MiB
& vs.\ dko
& MiB
& vs.\ dko
& MiB
& vs.\ dko
& time
& MiB
& vs.\ dko
& MiB
& vs.\ dko
\\
\bottomrule
\end{tabular}